\documentclass[10pt]{article}
\usepackage{a4}
\newcommand{\nn}{\nonumber}

\newcommand{\midd}[1]{\langle #1 \rangle}
\newcommand{\be}{\begin{equation}}
\newcommand{\ee}{\end{equation}}
\newcommand{\BE}{\begin{eqnarray}}
\newcommand{\EE}{\end{eqnarray}}
\newcommand{\ket}[1]{| #1 \rangle}
\newcommand{\bra}[1]{\langle #1 |}
\newcommand{\proj}[1]{| #1 \rangle \langle #1 |}
\newcommand{\inpr}[2]{\langle #1 | #2 \rangle}
\def\openone{\leavevmode\hbox{\normalsize1\kern-3.8pt\large1}}
\newcommand{\forget}[1]{}
\begin{document}
\title{
Interference and Distinguishability in Quantum Mechanics}
\date{April, 1988}
\author{Jos Uffink and Jan Hilgevoord}
\maketitle

\begin{abstract}
 Quantitative measures are introduced for the indistinguishability $U$ of
two quantum states in a given measurement and the amount of interference
$I$ observable in this measurement. It is shown that these measures obey
an inequality $U\geq I$ which can be seen as an exact formulation of
Bohr's claim that one cannot distinguish between two possible paths of a
particle while maintaining an interference phenomenon. This formulation is
applied to a neutron interferometer experiment of Badurek e.a. It is shown
that the formulation is stronger than an argument based on an uncertainty
relation for phase and photon number considered by these authors.
 \end{abstract}

A recent experiment in neutron interferometry \cite{badurek}
can be seen as a realisation of the double-slit  thought experiment
discussed by Einstein and Bohr. 
In this famous discussion, Bohr argued that one cannot distinguish
between two possible paths of a particle  while preserving 
an interference phenomenon.
To reach this conclusion Bohr applied the uncertainty relation for
position and momentum in a somewhat informal way. 
 However, it has been shown that the Heisenberg
uncertainty relation by itself is not strong enough to justify Bohr's
claim~\cite{UH}.

The discussion has been revived in the light of the new neutron
experiments.  In ref.~1 an argument is presented supporting Bohr's claim,
based on an uncertainty relation for the phase and photon number of the
electromagnetic field.  The validity of this explanation was subsequently
disputed \cite{dewdney}, because of the dubious theoretical status of the
phase-number uncertainty relations.

This raises the question whether it is possible to give a direct
quantitative formulation of Bohr's claim, without recourse to the
uncertainty relations. Work in this direction has been done by Wootters and
Zurek\cite{WZ}. Here we propose an alternative formulation that seems
particularly apt for the interferometer experiments. 

The main obstacle for a direct formulation of Bohr's claim is the
problem of choosing a quantitative measure for the extent to which the
two paths are distinguishable in a given experiment. Let $\psi_1$ and
$\psi_2$ denote orthogonal quantum states that represent possible paths
of a particle. A measurement performed on this particle may be described
by a complete set of orthogonal projection operators
$\{ D_k\}$, where $k$ denotes a possible outcome of the measurement.
The hypotheses 
that the particle traveled either one of two paths then provide two
probability distributions, viz.
\BE 
 p_k   &=& 
       \bra{\psi_1} D_k \ket{\psi_1}, \nn 
\\[-.5 \baselineskip]
&&  \\[-.5 \baselineskip]
 q_k   &=&  \bra{\psi_2} D_k \ket{\psi_2}. \nn\EE
The two paths may be said to
be discriminated if, from the observed outcome of the measurement one
can decide between these two hypotheses. Thus, the problem of
distinguishing between two paths can be seen as a special case of the
general classical problem of discriminating between two statistical
hypotheses. 

A solution to this problem depends, of course, on the distributions $p$
and $q$, but also on the observed outcome. 
However, independently of the  latter, one can 
indicate whether a discriminative answer is likely.   There are two
extreme cases. 
(i): $p_k q_k= 0 $  for all $k$, i.e.\ every outcome that has positive
probability according to one hypothesis is impossible according to the
other. 
In this case a single observation will suffice for complete
discrimination. (ii):   $p_k= q_k$ for all $k$. In this case, no number of
observations can discriminate between $p$ and $q$. In all other  cases
an incomplete discrimination is to be expected.

One is tempted to define a `degree of indistinguishability', reflecting
the expected lack of discrimination between $p$ and $q$.
For this purpose we choose 
 \be U(p,q)  = \sum_k \sqrt{p_k q_k}. \ee
The significance of this expression in statistical theory has been studies
by 
Bhattacharyya~\cite{Bhat}, Rao~\cite{Rao} and Wootters~\cite{Wootters}.
Loosely speaking, a value of $U(p,q)$
 close to unity indicates that even when one of these distributions is
`true' a typical outcome of the experiment will not allow one to infer
accurately which of them is true.

As an illustration we compare the expression proposed above with a more
familiar statistical approach to the discrimination problem, the theory 
of hypothesis testing~\cite{KS}. A `test' for two statistical hypotheses
is a procedure
by which
one
hypothesis is either accepted or rejected in favor of the alternative
hypothesis. 
In this approach the quality of the test
is judged by the so-called errors of the first and second kind,
$1-\alpha$ and $1 -\beta$; i.e., respectively, the probability of
rejecting the first
hypothesis when it is true, and the probability of accepting it when its
alternative is true. 
A Neyman-Pearson (NP) test'  is designed to minimize these two errors,
and always obeys   $(1-\alpha) + (1-\beta) \leq 1$.
One can show that for all NP tests  of the distributions
$p$ and $q$:
\BE (1- \alpha ) + (1 -\beta) &\geq& 1 - \sqrt{1- U^2},
\nn\\[-.5 \baselineskip]
 &&\\[-.5 \baselineskip]
 (1- \alpha )  (1 -\beta) &\leq& \frac{1}{4} U^2 \nn.\EE
Thus, a large degree of indistinguishability implies that the sum of the
two kinds of error in a NP test  is close to its upper bound 1, whereas a
small value of $U(p,q)$ implies that any NP test
 of $p$ and $q$ is `good'
in the sense that the product of the errors is small.

In agreement with (2) we define the degree of indistinguishability of the
quantum states $\psi_1$ and $\psi_2$  for the measurement $\{ D_k\}$
as 
\be U_{ \{D_k\} } (\psi_1, \psi_2) = \sum_k 
\sqrt{  
       \bra{\psi_1} D_k \ket{\psi_1}
       \bra{\psi_1} D_k \ket{\psi_1}}.
\ee
 Thus even when it is assumed that the particle did travel one of the
paths, the outcome of the measurement will only enable us to determine
this
path if $U$ is small.\footnote{%
	(Note added:) The meaning of the indistinguishability
        of two hypotheses is explained here in terms of the assumption
        that the particle does actually travel a definite path.
      One may object that, according to Bohr, this is not the case in a
     interference experiment. Indeed, in that case, the true state is not
	$\psi_1$ nor $\psi_2$ but a superposition of both.  However,
       the assumption is made here only for the purpose of
     elucidating  the meaning of `indistinguishability'.
         The point is that depending on whether
      $U_{\{D_k\}}(\psi_1,       \psi_2)$ is large (or small),
       the two	states are hard (or easy)  to tell apart. 
         Once this is accepted, the question whether $U_{\{D_k\}}(\psi_1,
	\psi_2)$ is indeed small or large is of course completely
      independent of the question whether $\psi_1$, $\psi_2$ or any other
     state is
      actually prepared in the experiment.}
One may easily show that for given $\psi_1$ and $\psi_2$
the
expressions (4) is non-decreasing when the projections are resolved into 
lower-dimensional projections, as is intuitively reasonable. Maximal
distinguishability is reached when the set 
$\{ D_k\}$ includes 
$\proj{\psi_1}$ or  
$\proj{\psi_2}$.

 We now turn to the notion of interference. 
The state of a particle emerging from an interferometer may be regarded as
a normalized superposition of $\psi_1$ and $\psi_2$, $\psi = 
c_1 \psi_1 +
c_2 \psi_2$,  with 
$|c_1|^2+
|c_2|^2 =1$. The probability that the measurement $\{D_k\}$ yields an
outcome $k$ is then 
 \be P_k = 
|c_1|^2 p_k + |c_2|^2q_k + i_k,\ee
where 
$p_k$ and $q_k$ are given by (1), and
\be
i_k = 2\, \mbox{Re}\, c_1^* c_2 
\bra{\psi_1}D_k \ket{\psi_2} \ee
is the so-called interference term.
 The maximum value of $i_k$ for all
choices of $c_1$ and $c_2$ is 
\[ I_k
= | 
\bra{\psi_1}D_k \ket{\psi_2} |.\]
We define the interference power of $\psi_1$ and $\psi_2$ for this
measurement as 
\be I_{ \{D_k\} } (\psi_1, \psi_2) = \sum_k  I_k  =
\sum_k |\bra{\psi_1}D_k \ket{\psi_2} |.\ee
Let us see how this relates to other familiar notions of interference
strength.   In many experiments a variable phase shift $\chi$ between the
two paths is introduced. In that case it is convenient
to put 
$\psi_2(\chi) =  e^{i\chi} \psi_2$.  Let us further take $c_1 = c_2  =
1/\sqrt{2}$. $P_k$ then oscillates as a function of $\chi$:
\[ P_k(\chi)  = \frac{1}{2}(p_k + q_k) +  |\bra{\psi_1} D_k\ket{\psi_2}|
 \cos \chi. \]
A well-known measure for the amount of interference in this situation is
the 
Michelson fringe visibility
\[ V_k = 
\frac{P_{\rm max} -P_{\rm min}}{
P_{\rm max} +P_{\rm min}},\]
where  $P_{\rm max}$ and $P_{\rm min}$ denote adjacent maximal and minimal
values of 
$P_k(\chi)$.
 This gives  \[ V_k = 2
 |\bra{\psi_1} D_k
\ket{\psi_2}|/ (p_k + q_k), \]or
\[
 I_{\{ D_k\}}
 = \sum_k  V_k \frac{1}{2}  (p_k + q_k).\]
Thus, the interference power is just the average visibility  over all
possible outcomes,
weighted by the mean probability $\frac{1}{2} (p_k + q_k)$.

For given $\psi_1$ and $\psi_2$,   $ I_{\{ D_k\}}$ is non-decreasing
when the projections are resolved into lower-dimensional projections, so
that
a more resolving measurement will in general show more interference.

Now, let us compare the interference power  (7) with the degree of
indistinguishability (4). 
By the Schwartz inequality, one finds  
\be
 U_{\{ D_k\}} 
(\psi_1, \psi_2)
 \geq I_{\{ D_k\}}(\psi_1, \psi_2).
\ee
Thus the appearance of a pronounced interference phenomenon in an
experiment is incompatible with  the requirement  that the interfering
states are distinguishable in that experiment. 
This can be regarded as a quantitative expression of Bohr's claim.
Equality in (8) occurs when all projections  $D_k$ are   one-dimensional. 

In relation  (8) the degree of indistinguishability and interference power
are compared for one and the same experiment. 
What if the interference is observed in one measurement, $\{ D_k\}$, and
one attempts to distinguish the states by means of another measurement, 
described by the set $\{ D'_l \}$? Actually this  will not improve the
situation, as long as the two measurements are compatible.   In that case,
the product of two projections, $  D_{kl} = D_k D'_l$, is also an
orthogonal projection, and  $D_{kl}$ is a resolution of 
$\{ D_k\}$ 
as well as $\{ D'_l\}$, so that 
\be    
U_{\{ D'_l\}} \geq 
U_{\{ D_{kl}\}}  \geq
I_{\{ D_{kl}\}}  \geq
I_{\{ D_k\}}.
\ee

We now apply the ideas discussed above to the neutron interferometer.
In the interferometer an incident neutron beam is coherently split 
in two partial beams, 1 and 2. A phase shift may be produced by placing a
piece of
material in one of them. Next the  two beams are again coherently split
and pairwise superposed, so that the neutrons emerge in two final
beams, $A$ and $B$. The emerging neutron state may then be written as a
superposition
of orthogonal parts corresponding to the two paths.
\be \phi^0=
\frac{1}{\sqrt{2}} (\phi^0 _1  + e^{i\chi} \phi^0_2),
\ee
where $\chi$ is the variable phase shift. In the ideal case, each of
the paths
 corresponds with equal amplitude to the emerging beams,
\BE
\phi^0_1 
&=&
\phi^0_{1A} 
+\phi^0_{1B}, \nn
 \\[-.5 \baselineskip]
&&\\[-.5 \baselineskip]
\phi^0_2 
&=&
\phi^0_{2A} 
+\phi^0_{2B}, \nn
\EE
and 
\be \phi^0_{1A} =  \phi^0_{2A},~~~~~\phi^0_{1B} =- \phi^0_{2B}. \ee
Thus, for $\chi=0$, all neutrons are found in beam 
$A$, whereas for a
phase shift $\chi=\pi$ all neutrons are found in beam $B$. This
conforms
to maximal interference as a function of the phase shift.

In the experiment of ref.~1 the incident beam is prepared with spin
polarized along the {$z$}-axis.  One of the paths is lead through
a magnetic coil which reverses the spin of the passing neutrons. In the
emerging beams spin analyzers and detectors are placed, and the 
time-dependent intensity is recorded. 
When the analyzers are turned to to analyze spin in the {$y$}-direction, a
periodical interference pattern in the intensity is observed as a function
of the phase shift.

The question is now whether one can infer, without disturbing the
interference, which path each neutron traveled through the interferometer
by means of an extra measurement on the spin flipping coil. The idea
behind this is that the interaction of the neutron with the magnetic field
of the coil involves the exchange of a photon. Hence if a measurement of
the photon number of this field is made, one might hope to detect whether
or not such an interaction has taken place.

Let the spin flipper be placed in path 2. The combined final state of
neutron and magnetic field  may be taken as 
\be
\psi =  \frac{1}{\sqrt{2}}
(
\psi_1
 + e^{i\chi}\psi_2
), 
\ee
with
  \be \psi_1
= \phi^0_1 \xi, \ee
where $\xi$ is the final state of the magnetic field when no interaction
occurs, and $\psi_2$ is the combined final state of neutron and
magnetic field when  the interaction  does take place.
Introducing the neutron eigenstates for the $z$-component of the spin,
 $\ket{\pm}$, and  photon number eigenstates $\ket{n}$, we may write:
\BE \phi^0_1 & = &\ket{+}\inpr{+}{\phi^0_1}, \\
\xi  & = &\sum_n   \ket{n} \inpr{n}{\xi} .\EE
The spin flipper is assumed to be efficient,
i.e.\ it fully reverses the spin of all passing neutrons.
Further, we assume that the interaction does not alter the spatial 
wave function of the neutron.   $\psi_2$ may then be written as 
\be \psi_2 = \phi_2 \xi_2, 
\ee
with 
\BE 
\phi_2  &= &  e^{i\omega t} \ket{-}\inpr{+}{\phi^0_2},  \\
\xi_2   & = & \sum_n   e^{-i\omega t} \ket{n+1}\inpr{n}{\xi},  \EE
where $\omega$ is the frequency of the electromagnetic field. 

Suppose now that a measurement is made of the intensity in the
emerging neutron beams with  spin analyzed in the $y$-direction at all
positions  $x$ along the beams at a fixed time.  It is easy to see that
this procedure, and the actual experiment, in which  a time-dependent
observation is made
at  a fixed  position, are equivalent in the quasi-monochromatic
approximation for the neutron wave packet. 

The interference power of the states  is (14) and (17) for this
experiment is:
\be
I_{ \{ D_{xs} \} } =  | \inpr{\xi}{\xi_2}| \sum_s \int\, dx 
|\bra{\phi^0_1} D_{xs} \ket{\phi_2} |,
\ee
where the integral is to be performed over both emerging beams. 
Using the fact that the projections  $D_{xs}$ can be factorized,
$ D_{xs} =   D_x D_s$, with 
\BE
D_{s= +\frac{1}{2}}  &=& 
 \big( \ket{+} + \ket{-} \big)
 \big( \bra{+} + \bra{-} \big), \nn\\
D_{s= -\frac{1}{2}}  &=& 
\big( \ket{+} - \ket{-} \big)
 \big( \bra{+} - \bra{-} \big). \nn\EE
We obtain from (11) and (12) 
\be
I_{ \{ D_{xs} \} } =  \left|\sum_n   \inpr{\xi}{n} \inpr{n+1}{
\xi}\right|.
\ee

Let us now ask whether the passage of a neutron through the spin flipper,
and the associated  photon exchange, can be detected by a measurement of
the photon number. The two states to be distinguished are (14) and (17). 
For the measurement of the photon number,
represented by the projections $D_n =\proj{n}$, the degree of
indistinguishability  (4) is 
\be 
U_{ \{ D_{n} \} } (\psi_1,\psi_2) =  \sum_n  \sqrt{|\inpr{\xi}{n}
\inpr{n+1}{ \xi}|}.
 \ee
Thus, the extent to which  one can determine whether a photon 
exchange has occurred, depends on the extent to which consecutive
pairs of photon numbers are represented in the distribution 
$ |\inpr{n}{\xi}|^2$. It is obvious from (21) and (22) that 
\be
U_{ \{ D_{n} \} } (\psi_1,\psi_2) 
\geq 
I_{ \{ D_{xs} \} } (\psi_1,\psi_2),  \ee
in agreement with (9).
This shows that the conditions under which
 a measurement on the spin
flipper 
allows the distinction of the two paths exclude the conditions under
which interference occurs. 

Finally, we compare the formulation of Bohr's claim given here with an
approach employing an uncertainty relation for the phase and photon
number. 
A satisfactory description of the phase of an electromagnetic fields in
quantum theory is that in terms of the exponential phase operator 
$e^{i\phi}$ defined by L\'evy-Leblond\cite{ll},\footnote{%
(Note added:)	The notation `$e^{i\phi}$' for this operator is merely
        formal and should not be
	mistaken as implying that it is 
	an  exponent of of some self-adjoint  operator $\phi$.
	In fact, $e^{i\phi}$ is not  even unitary.}
\[  e^{i\phi}  = \sum_n \ket{n}\bra{n+1}.
\]
Note that the expectation value of this operator appears in (21).
An appropriate definition for the phase uncertainty, as discussed in  
ref.~9, is \be
(\Delta \phi)^2 =  1 - |\bra{\xi}  e^{i\phi}  \ket{\xi} | -
|\inpr{0}{\xi}|^2 \ee
where $ |\inpr{0}{\xi}|^2 $ is the probability to find zero photons in the
magnetic field. Under experimental conditions this term may be
neglected. Therefore, in order to have an  appreciable interference
power (21),
 it is necessary for the  magnetic field that   $\Delta\phi \ll 1$.
We can now employ the uncertainty relation for phase and photon number
given 
in ref.~9:
\be
   (\Delta n)^2   \big( (\Delta \phi)^2 - \frac{1}{2} |\inpr{0}{\xi} |^2
\big) 
\geq \frac{1}{4} 
\big( 1 - \Delta \phi)^2 - |\inpr{0}{\xi} |^2 \big)
\ee
where 
$(\Delta n )^2 = \midd{n^2} - \midd{n}^2$ is the standard deviation of
the photon number in the state $\xi$. 
It then follows from condition (25) that  
\be \Delta n \gg \frac{1}{2}.\ee
However, this conclusion by itself is not sufficient to exclude  the
distinguishability of the two paths. 
For example, if the state of the photon field were such that 
\be
|\inpr{n}{\xi} |^2 = \frac{1}{2}   (\delta_{n n_0} + \delta_{nn_1}),~~~~
 | n_0 - n_1| \gg 1,
\ee
(where $\delta$ denotes the Kronecker delta)  
 the photon exchange could  still be detected with
complete certainty from a measurement of the photon number, 
without violating  (27).
However, the fact that such a state is inconsistent with condition (25)
follows directly from relation (23).

We conclude that the formulation of Bohr's claim considered above is
stronger than one based on an certainty relation of the Heisenberg type.

\end{document}